\documentclass[aps,twocolumn,longbibliography,english,superscriptaddress,prr]{revtex4-1}
\usepackage[colorlinks=true,urlcolor=blue,citecolor=blue,linkcolor=blue]{hyperref}
\usepackage[T1]{fontenc}
\usepackage{amssymb}
\usepackage{tabularx}
\usepackage[plain]{algorithm}
\usepackage{algpseudocode}
\usepackage{rotating}
\usepackage{booktabs}
\usepackage{xcolor}
\makeatletter
\newsavebox{\@brx}
\newcommand{\llangle}[1][]{\savebox{\@brx}{\(\m@th{#1\langle}\)}%
  \mathopen{\copy\@brx\kern-0.5\wd\@brx\usebox{\@brx}}}
\newcommand{\rrangle}[1][]{\savebox{\@brx}{\(\m@th{#1\rangle}\)}%
  \mathclose{\copy\@brx\kern-0.5\wd\@brx\usebox{\@brx}}}
\makeatother

\usepackage{bbm}
\usepackage{graphicx, subfigure}
\usepackage{amsmath,color}
\usepackage{mathrsfs}
\usepackage{float}
\usepackage[normalem]{ulem}
\usepackage{indentfirst}
\usepackage{txfonts}
\usepackage{qcircuit}

\emergencystretch=\maxdimen
\makeatletter

\def\journal #1, #2, #3, 1#4#5#6{{\sl #1~}{\bf #2}, #3 (1#4#5#6) }

\newcommand{\<}{\langle}
\renewcommand{\>}{\rangle}
\newcommand{\SWAP}{{\rm SWAP}}
\newcommand{\CNOT}{{\rm CNOT}}
\newcommand{\X}{{\rm X}}
\renewcommand{\H}{{\rm H}}

\newcommand{\SU}{{\rm SU}}
\newcommand{\UU}{{\rm U}}
\newcommand{\thetav}{{\boldsymbol{\theta}}}

\newcommand{\vtheta}{{\boldsymbol{\theta}}}

\newcommand{\sigmai}[2]{{\sigma^{#2}_{#1}}}
\newcommand{\qi}[1]{{q^{\alpha_{#1}}_{#1}}}

\newcommand{\circled}[1]{\raisebox{.5pt}{\textcircled{\raisebox{-.9pt} {#1}}}}

\newcommand{\expect}[2]{{\mathop{\mathbb{E}}\limits_{\substack{#2}}\left[#1\right]}}
\newcommand{\var}[2]{{\mathop{\mathrm{Var}}\limits_{\substack{#2}}\left(#1\right)}}

\newcommand{\Eq}[1]{Eq.~(\ref{#1})}
\newcommand{\Fig}[1]{Fig.~\ref{#1}}
\newcommand{\Tbl}[1]{Table~\ref{#1}}
\newcommand{\Ref}[1]{Ref.~\onlinecite{#1}}
\newcommand{\Sec}[1]{Sec.~\ref{#1}}
\newcommand{\App}[1]{Appendix \ref{#1}}
\newcommand{\bra}[1]{\mbox{$\left\langle #1 \right|$}}
\newcommand{\ket}[1]{\mbox{$\left| #1 \right\rangle$}}

\newcommand{\tr}[1]{\mathrm{tr}\mbox{$\left[ #1\right]$}}

\newcommand{\ra}[1]{\renewcommand{\arraystretch}{#1}}

\newcommand{\material}[1]{\iffalse[{\bf  \color{cyan}{Material: #1}}]\fi}

\makeatother

\usepackage{babel}

\begin{document}

\title{Variational Quantum Eigensolver with Fewer Qubits}

\author{Jin-Guo Liu}
\affiliation{Institute of Physics, Chinese Academy of Sciences, Beijing 100190, China}
\author{Yi-Hong Zhang}
\affiliation{Institute for Interdisciplinary Information Sciences, Tsinghua University, Beijing, 100084, China}
\author{Yuan Wan}
\affiliation{Institute of Physics, Chinese Academy of Sciences, Beijing 100190, China}
\author{Lei Wang}
\email{wanglei@iphy.ac.cn}
\affiliation{Institute of Physics, Chinese Academy of Sciences, Beijing 100190, China}
\affiliation{CAS Center for Excellence in Topological Quantum Computation, University of Chinese Academy of Sciences, Beijing 100190, China}
\affiliation{Songshan Lake Materials Laboratory, Dongguan, Guangdong 523808, China}

\begin{abstract}
    We propose a qubit efficient scheme to study ground state properties of quantum many-body systems on near-term noisy intermediate scale quantum computers. One can obtain a tensor network representation of the ground state using a number of qubits smaller than the physical degrees of freedom. By increasing the qubits number, one can exponentially increase the bond dimension of the tensor network variational ansatz on a quantum computer. Moreover, we construct circuits blocks which respect $\UU(1)$ and $\SU(2)$ symmetries of the physical system and show that they can significantly speed up the training process and alleviate the gradient vanishing problem. To demonstrate the feasibility of the qubit efficient variational quantum eigensolver in a practical setting, we perform first principle classical simulation of  differentiable programming of the circuits. Using only $6$ qubits one can obtain the ground state of a $4\times 4$ square lattice frustrated Heisenberg model with fidelity over 97\%. 
Arbitrarily long ranged correlations can also be measured on the same circuit after variational optimization. 
\end{abstract}
\maketitle

\section{Introduction}
Studying ground state properties of quantum many-body systems is a promising native application of quantum computers. Given limited qubit resources and noisy realizations of near-term quantum  devices~\cite{Preskill2018, Boixo2018}, a practical approach is to employ the variational quantum eigensolver (VQE)~\cite{Peruzzo2014, Jarrod2016, Wecker2015a, Wecker2015b, McArdle2018, Cao2018}, which runs in a classical-quantum hybrid mode.
In this scheme, a parametrized quantum circuit provides a variational ansatz for the ground state. A classical optimizer tunes the circuit parameters to reduce the expected energy of the target Hamiltonian of the output quantum state. There were already several small scale experimental demonstrations of VQE for molecules and quantum magnets~\cite{Shen2017, OMalley2016, Kandala2017,Colless2018, Hempel2018}. These early experiments mostly employed gradient free or Bayesian approaches for classical optimization. Recent progress on unbiased gradient estimation on quantum circuits~\cite{Li2017a, Mitarai2018, Liu2018, Verdon2018, Schuld2018, Javier2018, Bergholm2018, Guerreschi2017,Farhi2018,Romero2018,Harrow2019,Dallaire2018} breaks the information bottleneck between classical and quantum processors, thus
providing a route towards scalable optimization of circuits with a large number of parameters.

There are nevertheless more challenges in the training of variational quantum circuits.
The gradients of an unstructured, randomly parametrized circuit vanish exponentially as a function of the number of parameters~\cite{McClean2018} due to the concentration of measure in high dimensional spaces ~\cite{Gross2009,Bremner2009}.
Intuitively, this could be understood by the fact that the overlap between a random initial
quantum state and a target state is exponentially small in the many-body Hilbert space. This difficulty motivates one to design the circuit architecture and initialize the circuit parameters with insights from classical tensor networks~\cite{Huggins2019, Kim2017,Peng2019}
and quantum chemistry ansatz~\cite{Yung2014, Lee2018, Barkoutsos2018}. 
Furthermore, since the number of required qubits is the same as the problem size in the standard VQE applications, one has to push up the number of controllable qubits way beyond the current technology to convincingly surpass the classical simulation approach in finding the ground states of quantum many-body systems.
Related approaches such as  the quantum approximate optimization algorithm~\cite{Farhi2014} and related field such as quantum machine learning~\cite{Ciliberto2017, Mitarai2018, Liu2018} suffer from the same problem.

We address these problems by adopting the qubit efficient circuit architecture~\cite{Monras2010, Huggins2019} for the variational quantum eigensolver. By measuring qubits sequentially and reusing the measured qubit, one can produce quantum states for an arbitrarily large system with a fixed number of qubits. This approach amounts to generating matrix product states (MPS)~\cite{Ostlund1995, Vidal2003, Schollwock2011} on a quantum computer~\cite{Schon2005, Cramer2010}. Note that, although it is well known that an MPS with small bond dimension can be efficiently simulated classically~\cite{Vidal2003}, having quantum resource allows one to reach exponentially large bond dimension that is inaccessible to classical computers.
Despite its one-dimensional geometry, MPS is a versatile variational ansatz that has been successfully applied to systems with diverse lattice geometry and topology~\cite{Stoudenmire2012}. The success lies in the fact that many physics and chemistry systems of interest exhibit relatively small entanglement entropy in their ground states~\cite{Verstraete2004, Eisert2010}.

A natural extension of MPS to higher dimensions is projected entangled pair states (PEPS)~\cite{Verstraete2006}, which is able to represent many area law entangled two dimensional quantum states with only polynomial number of parameters. However, classical algorithm for exact contraction of a finite size PEPS shows exponentially complexity with the problem size. 
We propose a PEPS-inspired quantum circuit ansatz for the ground states of two-dimensional quantum systems. The ansatz can capture area law entanglement entropy. Although classical simulation of this ansatz is exponential difficult, one can estimate the unbiased energy expectation of the ansatz on a quantum computer efficiently. 

\begin{figure}
    \begin{center}
        \includegraphics[width=\columnwidth,trim={1cm 2cm 0.5cm 2cm},clip]{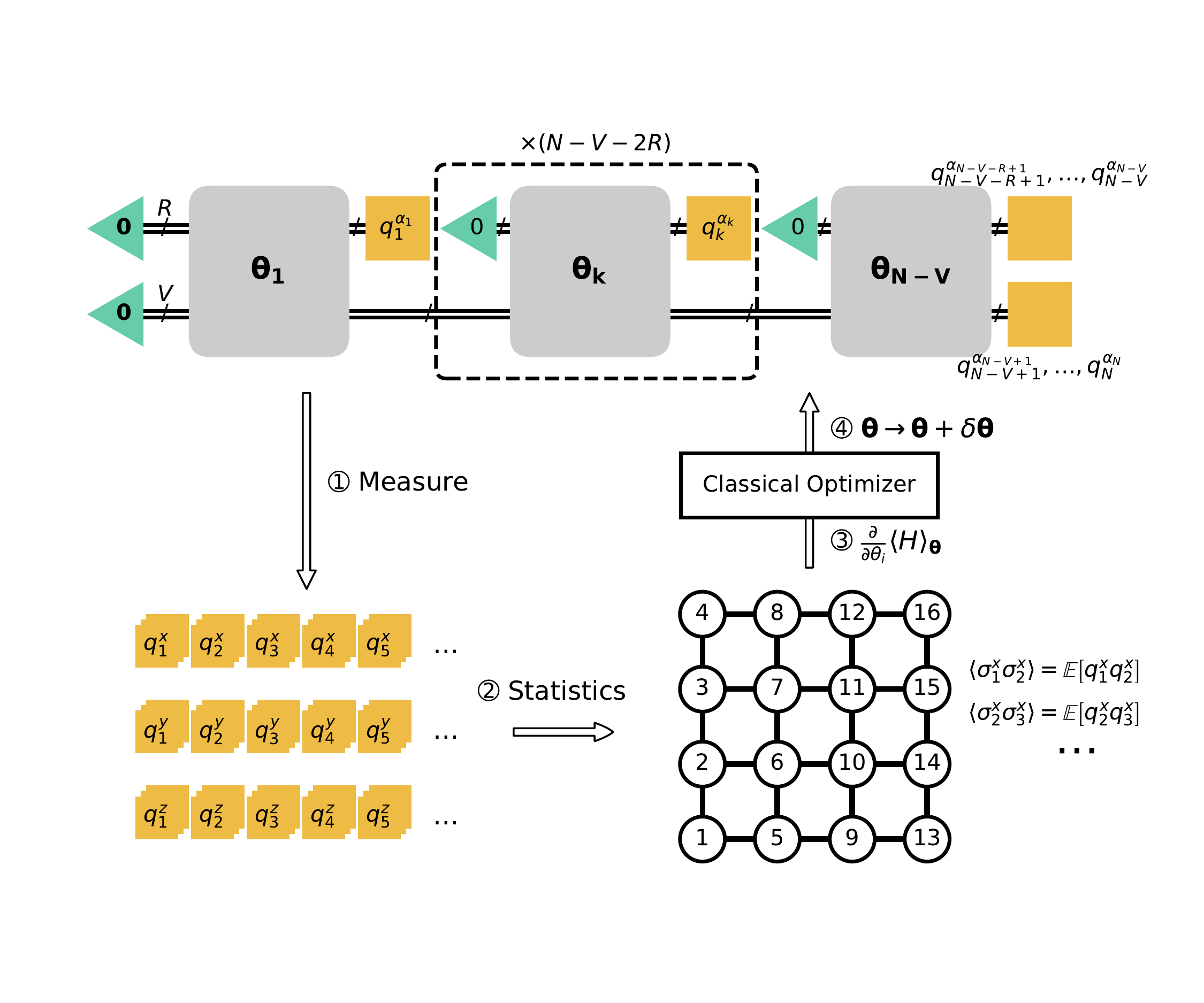}
    \end{center}
    \caption{Variational training of a MPS/PEPS prepared on a parametrized quantum circuit in the qubit efficient scheme. The upper left is the quantum circuit with $R$ reusable qubit for the physical degrees of freedom and $V$ qubits for the virtual degrees of freedom. Green triangles represent input single qubit state initialized to $\ket{0}$ and the yellow square $\qi{k}$ is the $k$-th output bit measured on Pauli basis $\alpha_k$. After measurement, the first qubit is reset to $|0\>$ and  then entangled with the remaining $V$ qubits before the next measurement. Each gray box represents a multi-layer parametrized quantum circuit detailed in \Fig{fig-circuits}. Each block has a similar circuit structure and the circuit parameters $\vtheta_1,\vtheta_2\ldots\vtheta_{N-V}$ are independent.}\label{fig-framework}
\end{figure}



Variational optimization of MPS generated on an actual quantum device has previously been demonstrated in an experiment~\cite{Eichler2015}. The experiment exploits the fact that the radiation field of a cavity QED naturally realizes~\cite{Barrett2013} the continuous MPS~\cite{Verstraete2010} with a few tuning parameters. Here, we focus on variational MPS and PEPS calculation on programmable gate model quantum computers.
This setup provides us with more systematic and precise control of the bond dimension and the number of variational parameters. Moreover, crucial technical advances such as gradient-based learning~\cite{Li2017a, Mitarai2018, Liu2018, Verdon2018, Schuld2018, Javier2018, Bergholm2018, Guerreschi2017,Farhi2018,Romero2018,Harrow2019,Dallaire2018} and quantum number preserving circuit design greatly speed up the training process, and make it practically useful for solving challenging quantum many-body problems.

This paper is organized as follows. In \Sec{sec-model} we introduce qubit efficient scheme of preparing MPS/PEPS on quantum circuits and a gradient-based variational training approach for obtaining the ground state of generic quantum many-body Hamiltonians. In \Sec{sec-example}, we demonstrate the utility of this scheme by numerically simulating the VQE of a frustrated Heisenberg model using fewer qubits than the system size. In \Sec{sec-gatecount}, we carry out a detailed gate counting to estimate the timings of actual experiments, followed by a discussion in \Sec{sec-discussion} which point to future research directions. Codes and pretrained circuits parameters can be found at the Github repository~\cite{github}.

\section{Circuit architecture and training approach}\label{sec-model}

Considering a generic quantum many-body Hamiltonian written in terms of the Pauli operators
\begin{equation}
    H = \sum\limits_{i\alpha} h_\alpha^i \sigmai i \alpha + \sum\limits_{ij\alpha\beta} h_{\alpha\beta}^{ij} \sigmai i\alpha\sigmai j\beta + \ldots,
\end{equation}
where $i,j=1,2\ldots N$ are site indices with $N$ being the system size, $\alpha,\beta = x, y, z$ are indices of the Pauli axis. We only show the first few terms for the sake of concreteness although higher order polynomials of Pauli operators are allowed.
In the variational quantum eigensolver approach~\cite{Peruzzo2014}, the ground state $|\psi(\vtheta)\>$ is represented by the output of a parametrized quantum circuit, where  $\vtheta = \left\{\theta_i\right\}$ are the circuit parameters. The variational energy is a summation of expectation values of Hamiltonian terms in the variational ground state,
\begin{equation}
    \langle {H} \rangle_\vtheta = \sum\limits_{i\alpha} h_\alpha^i \<\psi(\vtheta)|\sigmai i\alpha|\psi(\vtheta)\> + \sum\limits_{ij\alpha\beta} h_{\alpha\beta}^{ij} \<\psi(\vtheta)|\sigmai i\alpha\sigmai j\beta|\psi(\vtheta)\> + \ldots
\end{equation}
To estimate the expected energy, one can identify maximally commuting sets of Hamiltonian operators and measure all the commuting terms together on the corresponding bases.

One can prepare an MPS/PEPS as a variational state of $N$ qubits using a smaller number $R+V \ll N$ of qubits~\cite{Schon2005, Cramer2010, Huggins2019}, which will be dubbed Q-MPS/Q-PEPS hence force. The idea is to treat $R$ (for Q-MPS, $R=1$) of them as the physical qubit and use the remaining $V$ qubits as virtual degrees of freedom to mediate quantum entanglement. Sequentially measuring and reusing the physical qubits allow for producing an arbitrarily long MPS/PEPS. The circuit structure is illustrated schematically in the upper left corner of \Fig{fig-framework}. First, one initializes the $R+V$ qubits to the product state $\ket{\mathbf{0}}_R\otimes\ket{\mathbf{0}}_V$ and applies a circuit block parametrized by $\vtheta_1$ to all qubits. Then, one measures the physical qubits on Pauli basis $\sigmai{1}{\alpha_1}$ and stores the output $\qi{1}$ to a classical memory. Next, one recycles the measured qubit and reset it to state $\ket{\mathbf{0}}_R$. One then entangles them with the remaining $V$ qubits again by applying a second circuit block with parameters $\vtheta_2$. After repeating these procedures until one has collected $N-(R+V)$ bits of classical information, one measures all qubits to collect the last $R+V$ bits. This sequential measure-and-reuse scheme is equivalent to sampling this $N$-qubit MPS/PEPS on the same basis. Comparing to the qubit saving scheme for generative modeling in \Ref{Huggins2019}, here one to perform measurement on multiple different bases for the VQE calculation. 

The proposed scheme directly applies to Hamiltonians with arbitrarily long-range interaction. In particular, fermionic systems can be easily studied using the Jordan-Wigner transformation~\cite{Ortiz2001}. A general quantum chemistry problem is more challenging than the quantum spin problem considered here since they contain $\mathcal{O}(N^4)$ terms in the Hamiltonian. Nevertheless, the total number of measurements can be reduced using the techniques of~\cite{Cao2018, Babbush2018, Motta2018}. In \App{app-cluster}, we provide a concrete example of preparing and sampling a cluster state in the qubit efficient scheme.


Given the variational circuit, we aim at solving the optimization problem $\vtheta_{\rm opt} = {\rm argmin}_{\vtheta}   \langle {H} \rangle_\vtheta$. Gradient-based optimization algorithms are crucial to scaling to a large number of variational parameters~\cite{Liu2018}. Suppose that all the parameters of the quantum circuit appear in the form $e^{-i\theta_i\Sigma/2}$ with $\Sigma^2 = 1$. The analytical expression of the gradient with respect to the parameter $\theta_i$ reads~\cite{Mitarai2018},
\begin{equation}\label{eq-grad}
    \frac \partial{\partial \theta_i}\<H\>_{\vtheta} = \frac 1 2\left( \langle {H} \rangle_{\vtheta+ \frac \pi 2 \boldsymbol{e}_i} - \langle {H} \rangle_{\vtheta- \frac \pi 2 \boldsymbol{e}_i}\right).
\end{equation}
One can thus estimate the energy gradient by tuning the parameters to $\theta_i \pm \pi/2$ and use it for   gradient descend optimization of the energy. Unlike numerical differentiation, Eq.~(\ref{eq-grad}) is an exact gradient estimator, which is crucial for unbiased stochastic optimization with a noisy estimate of the gradients~\cite{Tucker2017b}.

To recapitulate, the key point of the proposed variational algorithm is to estimate the energy gradient of an $N$-qubit Hamiltonian with respect to an Q-MPS/Q-PEPS with fewer $(R+V\ll N)$ qubits. The steps are shown in \Fig{fig-framework}. $\circled{1}$ Tune a selected circuit parameter $\theta_i$ to $\theta_i +\frac \pi 2$, and collect bit strings by repeated measurements on various bases according to the Hamiltonian terms. Then we repeat for $\theta_i-\frac \pi 2$.
\circled{2}
Estimate the energy expectation value by assembling the statistics of all Hamiltonian terms
$\circled{3}$ Estimate gradient of all parameters via \Eq{eq-grad}. Feed the gradient information into a classical optimizer. $\circled{4}$ Update the circuit parameters according to suggestions of the classical optimizer. This completes one training epoch. The training stops when a prescribed convergence criterion is met. After reaching convergence, one may measure physical observables of interests on the optimized circuits.

\section{Application to Heisenberg model}\label{sec-example}
As a concrete example, we apply the approach detailed in the previous section to the frustrated Heisenberg model on a square lattice:
\begin{align}
    \begin{split}
        H = &\frac 1 4 \left[\sum\limits_{\<i,j\>}\sigmai{i}{x} \sigmai{j}{x}+\sigmai{i}{y}\sigmai{j}{y}+\sigmai{i}{z}\sigmai{j}{z}\right. \\
        &\left.+ J_2\sum\limits_{{\footnotesize\llangle} i,j{\footnotesize\rrangle}}\sigmai{i}{x} \sigmai{j}{x}+\sigmai{i}{y}\sigmai{j}{y}+\sigmai{i}{z}\sigmai{j}{z}\right],
    \end{split}
    \label{eq:heisenberg}
\end{align}
where $\<i,j \>$ and $\llangle i,j\rrangle$ denote nearest and next-nearest neighbors pairs, respectively. $J_2>0$ is the strength of the frustration term that suppresses the Neel order. The energy expectation value and its gradient can be efficiently evaluated by sampling the circuit output on three bases $\sigma^x$, $\sigma^y$ and $\sigma^z$.
In the following discussion, we consider the model on an open square lattice of the size $N=4\times 4$ with $J_2=0.5$. These sites are zigzag ordered in our ansatz as shown in the lower right of \Fig{fig-framework}.

Frustrated quantum spin models are crucial to the study of quantum magnets with many open problems~\cite{Zhou2017, Lucile2017}. Classical computational approaches to these problems are either limited by the sign problem~\cite{Troyer2005} or high computational cost at larger bond dimensions~\cite{Roman2018}. Variational optimization of Q-MPS/Q-PEPS on near-term quantum computers is a promising approach which may deliver valuable insights into open problems in this field.

\subsection{MPS inspired ansatz with conserved quantum numbers}
\begin{figure}
    \begin{center}
        \includegraphics[width=\columnwidth,trim={0cm 0.5cm 0.5cm 0cm},clip]{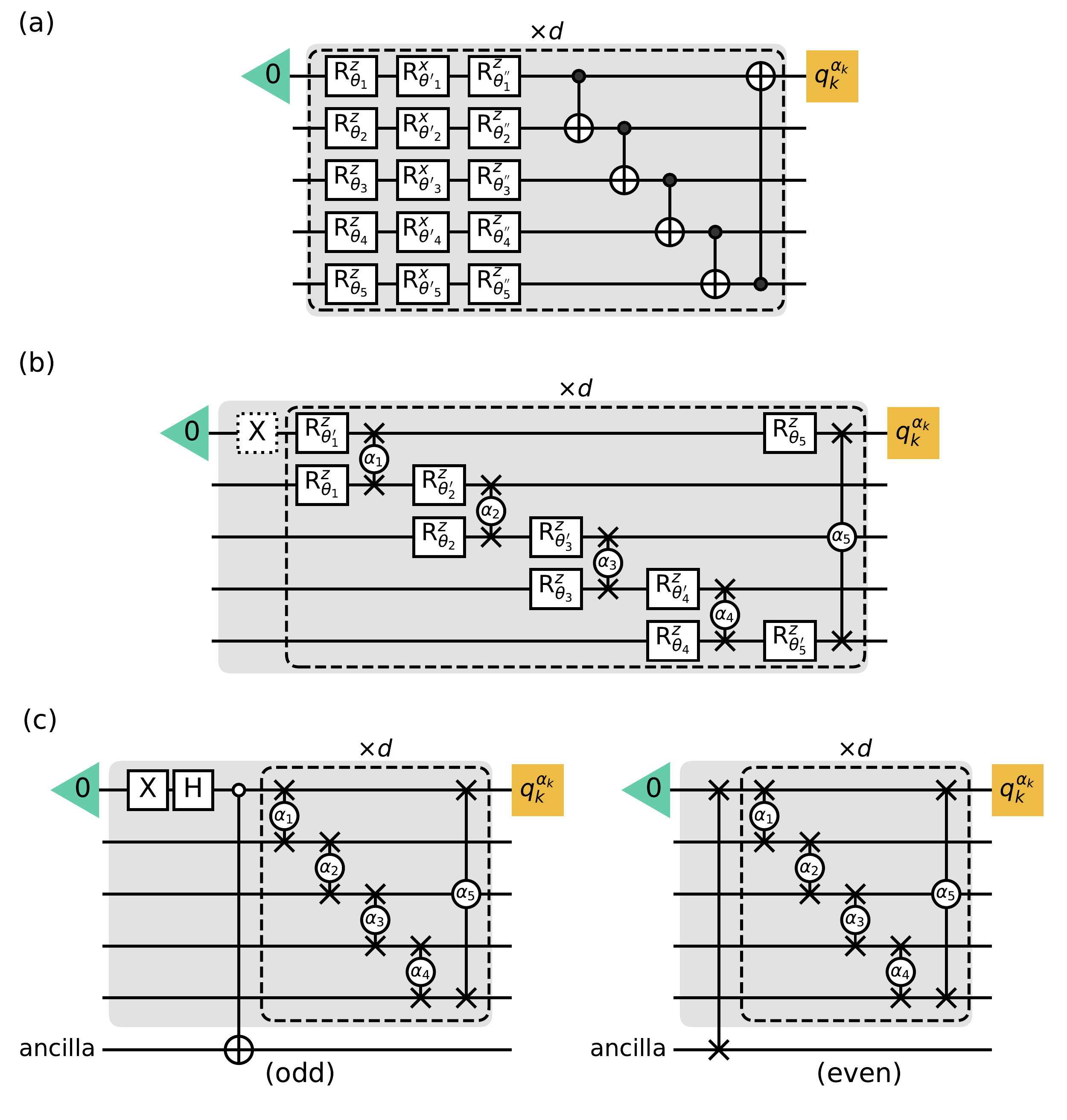}
    \end{center}
    \caption{The internal structure of circuit blocks shown in \Fig{fig-framework}.
    (a) a general unstructured setup.  $R^{\alpha}_{\theta_i}=e^{-\theta_i\sigmai{i}{\alpha}/2}$ represents a parametrized single qubit rotation gate.
    (b) $\UU(1)$ preserving block. The leftmost X gate is applied only for odd steps to flip the input state to $|1\>$. The double crosses are the $\SWAP^\alpha$ gates~\cite{Williams2011}.
    (c) $\SU(2)$ preserving block. The left and right panels are for odd and even steps respectively. The last qubit is an ancilla for creating singlets between consecutive steps. The gates enclosed in the dashed box are repeated for $d$ times, where $d$ is denoted as the depth of the block.}
    \label{fig-circuits}
\end{figure}

Figure~\ref{fig-circuits}(a) shows a general internal structure of the variational circuit which is efficient to be implemented on quantum hardware~\cite{Kandala2017}. Each layer contains $3(V+1)$ parameters in the rotational gates $R^x_\theta$ and $R^z_{\theta}$. We use $\CNOT$ gates with no variational parameters as the entanglers to generate entanglement between qubits. We repeat this construction for $d$ times within each circuit block. Thus there are $M=3d(V+1)$ parameters in each block. 
As we show below, taking into account the physical symmetries in designing of the VQE ansatz can reduce the number of parameters and increase the training performance.

\begin{figure*}
    \begin{center}
        \includegraphics[width=\textwidth,trim={3cm 9cm 3cm 8cm},clip]{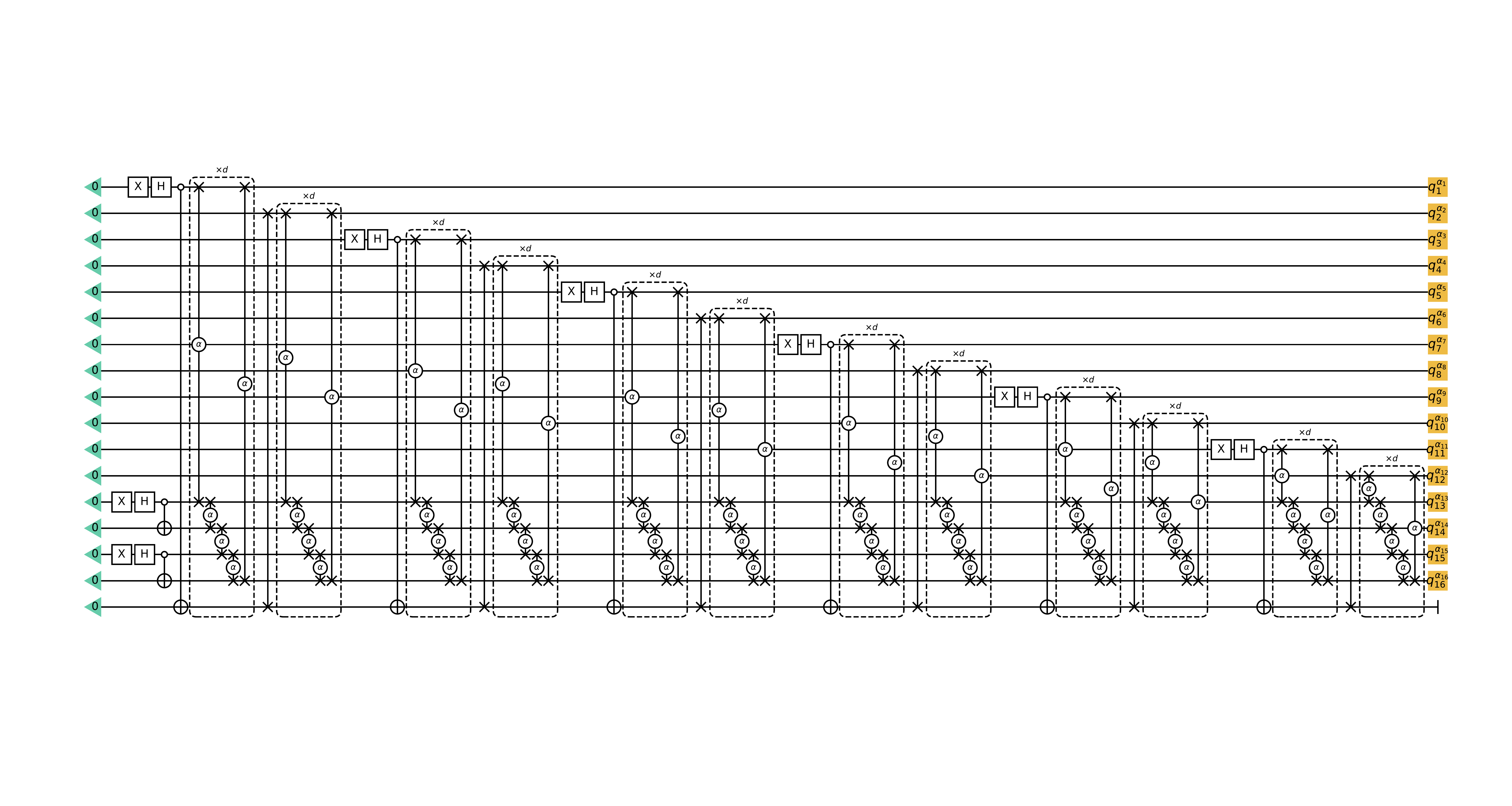}
    \end{center}
    \caption{Expanded view of the $\SU(2)$ symmetric quantum circuit ansatz shown in \Fig{fig-circuits}(c) for the $N=4\times 4$ Heisenberg model. The blocks inside a dashed box are repeated by $d$ times. In the qubit efficient scheme, the $1-12$-th qubits are the same physical qubit which is reused after measurement. The $13-16$-th qubits are the $V=4$ qubits which mediate entanglement in the final output state. The $17$-th qubit is the ancilla qubit for constructing singlet product initial state. The operations for generating singlets (i.e. X, H, controlled gates and SWAP gates) commute with parametrized swap gates in the dashed box, so that can be moved to the beginning of the circuit.}
    \label{fig-su2}
\end{figure*}

The Heisenberg model (\ref{eq:heisenberg}) has a $\UU(1)$ symmetry with good quantum number $S_z$. To preserve this symmetry~\cite{Jiang2018, Barkoutsos2018}, we construct a circuit block consists of $e^{-i\theta \sigma^z_i/2}$ and $e^{-i\theta\SWAP(i,j)/2}$ gates~\footnote{We note that these two gates are not sufficient to represent all $\UU(1)$ symmetric operations, e.g. the CZ gate. A general $\UU(1)$ ansatz which covers the whole symmetry sector will need all generators such as $\prod\limits_i \sigma^{z}_{i}$ and $\prod\limits_{ij} \SWAP(i,j)$, where $i$ runs over a subset of qubit indices.}. The latter gate is equivalent up to a phase factor to the $\SWAP^\alpha$ gate with $\alpha = \theta/\pi$~\cite{Williams2011}. Viewing the setup as a wide circuit with $N$ qubits, it is clear that the quantum number of the initial state is conserved during the evolution. To obtain the ground state in $S_z = 0$ sector, we prepare a spin-balanced initial state for the variational calculation. \Fig{fig-circuits} (b) shows that, by applying an additional X gate before the variational gates in the odd steps, one has an anti-ferromagnetic product state $|1010\ldots10\>$ as the initial state.

One can further exploit the full $\SU(2)$ symmetry of the Heisenberg model (\ref{eq:heisenberg}).
While there are sophisticated approaches~\cite{Weichselbaum2012,Schmoll2018} to implement this non-Abelian symmetry in classical simulations, the implementation is straightforward on quantum circuits. As shown in \Fig{fig-circuits} (c), we first prepare the input state in the total spin $\mathbf{S}^2=0$ sector, where the simplest choice is the singlet product state $|\psi_0\>=\bigotimes\limits_{i=1}^{N/2} |\uparrow\downarrow-\downarrow\uparrow\>$.
To prepare $|\psi_0\>$, we use an additional ancilla qubit to carry the entanglement of the physical qubit in the odd and even steps~\footnote{The ancilla qubit increases the bond dimension of the MPS to $2^{V+1}$.}. In the odd step, we prepare a spin singlet between the physical qubit and the ancilla qubit $|{\uparrow\downarrow-\downarrow\uparrow}\> = \overline{\CNOT(1, {\rm a})} \H(1) \X(1)|0\>_1\otimes|0\>_{\rm a}$. Here, $\overline{\CNOT(1, {\rm a})}$ is the inverse controlled-NOT gate, which flips the ancilla qubit when the physical qubit is in state $|0\>$. In the even step, we swap the ancilla and physical qubits. The physical qubits in the odd and even steps thus form a spin singlet. Then, we repeatedly apply parametrized $\SU(2)$ symmetric operations to the initial state to generate the variational output. We choose the generators to be the $\SWAP^\alpha$ gate~\cite{Williams2011} between a collection of qubits pairs~\footnote{We note that the variational ansatz may have a connection to the Bethe Ansatz~\cite{Murg2012} construction of the Q-MPS, where the scattering matrix has the same form as the parametrized SWAP gates.}.
Figure~\ref{fig-su2} shows the expanded view of the $\SU(2)$ symmetric quantum circuit. Parallel measurement of the final quantum state yields the same outcomes as the sequential measurement in the qubit efficient scheme shown in \Fig{fig-circuits}(c).

The $\SU(2)$ symmetric variational state reads $|\psi(\vtheta)\>=\prod\limits_{\{i,j\}} e^{-i\theta_{i,j}\SWAP(i,j)/2}|\psi_0\>$, where the product is ordered by the circuit architecture. This variational ansatz resembles the classical variational ansatz for quantum spins in the valence bond basis~\cite{Liang1988}. However, in general, the state could not be sampled efficiently using the classical Monte Carlo method due to the appearance of complex weights~\cite{Lou2007}. Moreover, since the swap operations are not commuting within each other, there is an additional difficulty in devising an efficient classical Monte Carlo scheme to sample from the variational ansatz. Therefore, variational optimization of this ansatz on a quantum device highlights the possible quantum advantage of the proposed qubit efficient VQE scheme.

To assess the feasibility of the qubit efficient VQE scheme on near-term quantum devices, we perform a faithful classical simulation of the training process. We simulate a circuit of $V+1$ qubits instead of an equivalent $N$ qubits circuit without qubit reusing. Therefore, even in the classical simulation, we do not have direct access to the final wavefunction but only to the measured bit strings. We sample the energy and its gradient on samples of batch size $4096$. Note that we purposely do not exploit the classical backpropagation algorithm (which reduces the complexity of gradient estimation from $\mathcal{O}(M^2)$ to $\mathcal{O}(M)$, $M$ is the number of variational parameters) to be in line with the repetitive experimental measurement~\cite{Liu2018}.

We use $V=4$ qubits for the virtual degrees of the MPS. The maximum entanglement entropy of the ansatz is thus $4 \ln 2$ given the full capacity of the variational blocks. We employ the Adam optimizer with a learning rate $0.1$~\cite{Kingma2014} for the stochastic gradient descent training. We compared three different circuit blocks shown in \Fig{fig-circuits}.
All of them have fixed depth $d=5$. The variational parameters are random initialized with uniform distribution in $[0,\pi]$. As shown in Fig.~\ref{fig-framework}, there are in total $N-V = 12$ circuit blocks.

The general circuit structure shown in \Fig{fig-circuits}(a) contains $M=3(V+1)(N-V)d=900$ variational parameters. In $500$ steps of training, the energy per site decreases to $-0.416$. 
For comparison, the exact ground state energy per site is $E_\mathrm{exact} = -0.46909731$, the density matrix renormalization group result for bond dimension $2^V = 16$ is $E_\mathrm{DMRG} = -0.46532670$~\footnote{For bond dimension $32$, $E_\mathrm{DMRG} = -0.46868064$, which represents the entanglement upper bond for the $\SU(2)$ symmetry preserving ansatz}. As the energy decreases, its fidelity with respect to the exact ground state increases from $5.7\times10^{-3}$ to $0.69$.

\begin{figure}
    \begin{center}
        \includegraphics[width=\columnwidth,trim={0cm 0.3cm 0cm 0.3cm},clip]{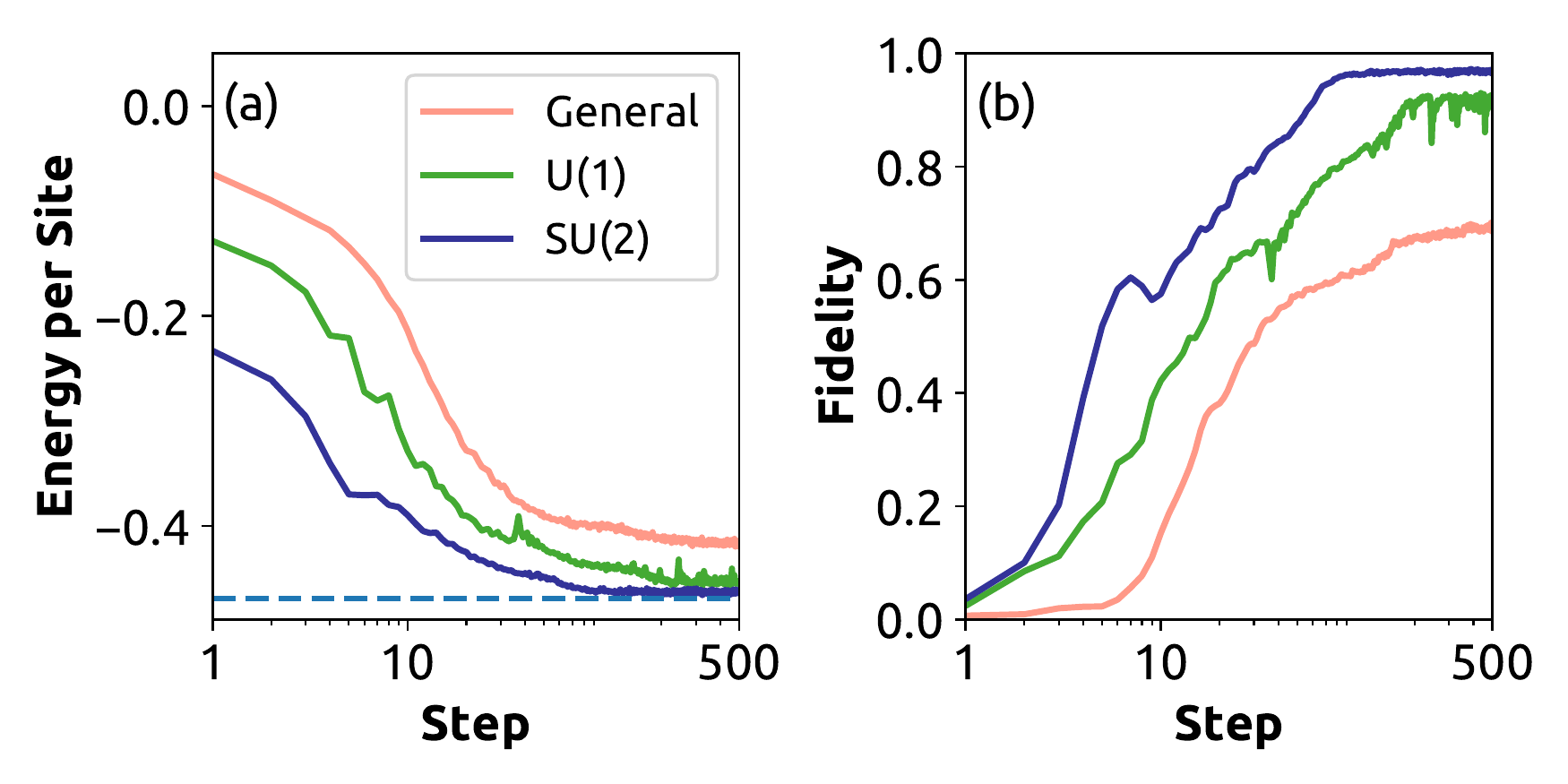}
    \end{center}
    \caption{(a) Variational energy and (b) fidelity with respect to the exact ground state of a $4\times4$ frustrated Heisenberg model as a function of gradient descent steps. The dashed line is the exact ground state energy. The circuit consists of $V+1=5$ qubits. For the $\SU(2)$ ansatz there is one  additional ancilla qubit, see \Fig{fig-circuits}(c). }
    \label{fig-eng-fid}
\end{figure}

Next, the $\UU(1)$ symmetric circuit structure in \Fig{fig-circuits}(b)
contains $10$ single qubit gates and $5$ two-qubit gates in each layer. Hence the number of circuit parameters is also $900$. However, the training efficiency increases significantly as shown in Figure~\ref{fig-eng-fid}. The ground state energy per site reaches $-0.454$, with a ground state fidelity of $0.92$.

\begin{figure}
    \begin{center}
        \includegraphics[width=\columnwidth,trim={0.3cm 0.35cm 0cm 0cm},clip]{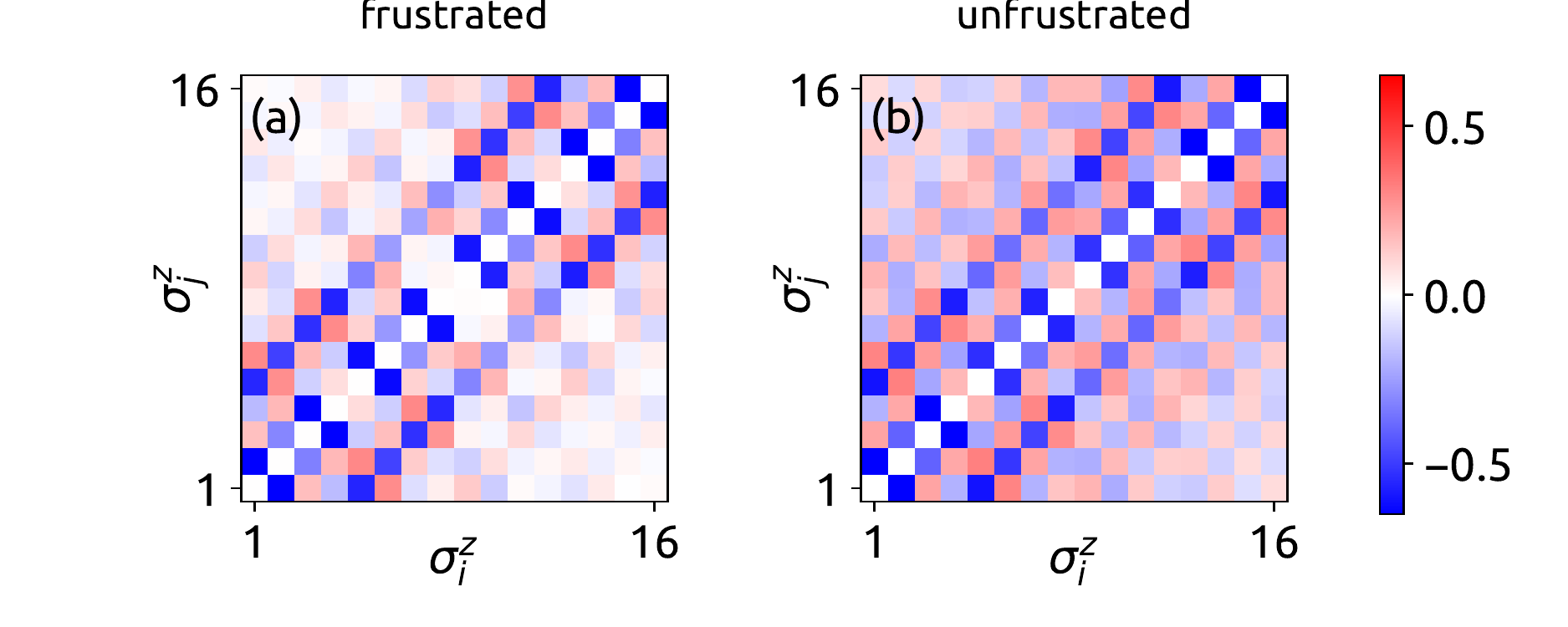}
    \end{center}
    \caption{Spin-spin correlation $\< \sigma_i^z \sigma_j^z\>$ for (a) frustrated Heisenberg model with $J_2=0.5$ and (b) unfrustrated Heisenberg model with $J_2=0$. Here, the ground state is obtained by using the $\SU(2)$ symmetric circuit shown in \Fig{fig-circuits}(c).}
    \label{fig-correlation}
\end{figure}

Finally, the $\SU(2)$ symmetric circuit structure in \Fig{fig-circuits}(c) gives the best variational energy despite that it only has $M=(V+1)(N-V)d=300$ variational parameters.
Using the same hyperparameters for training, the variational energy decreases to $-0.463$, and the fidelity reaches $0.97$. After obtaining the variational state, we can measure physical observables on the circuit. For example, the spin-spin correlation in $z$ direction $\<\sigmai i z\sigmai j z\>$ measured on the $\SU(2)$ symmetry preserving circuit shown in \Fig{fig-correlation} (a). As a comparison, using the same training hyperparameters we obtain a ground state with fidelity $0.98$ using $\SU(2)$ symmetric variational circuit ansatz \Fig{fig-circuits}(c) for unfrustrated Heisenberg lattice with $J_2=0$. As shown in \Fig{fig-correlation} (b), the checkerboard pattern for the antiferromagnetic correlation is more visible in the unfrustrated case.
These results suggest that, even with a moderate number of qubits, the qubit efficient VQE scheme is able to offer useful physical insights.

This result benefits from the fact that Q-MPS structure alleviates the gradient vanishing problem for studying low entangle physical systems.
The gradients of an unstructured, randomly parametrized circuit vanish exponentially as a function of the number of parameters~\cite{McClean2018} due to the concentration of measure in high dimensional spaces ~\cite{Gross2009,Bremner2009}.
Intuitively, this could be understood by the fact that the overlap between a random initial quantum state and a target state is exponentially small in the many-body Hilbert space. Related approaches such as  the quantum approximate optimization algorithm~\cite{Farhi2014} and related field such as quantum machine learning~\cite{Ciliberto2017, Mitarai2018, Liu2018} suffer from the same problem.

We inspect the variance of the gradient signal for various system sizes to investigate the gradient vanishing problem in the training variational quantum circuits~\cite{McClean2018}. To compute the gradient variance, we sample $1000$ gradients of random circuit parameters.

\begin{figure}
    \begin{center}
        \includegraphics[width=0.8\columnwidth,trim={0.3cm 0.5cm 0.4cm 0cm},clip]{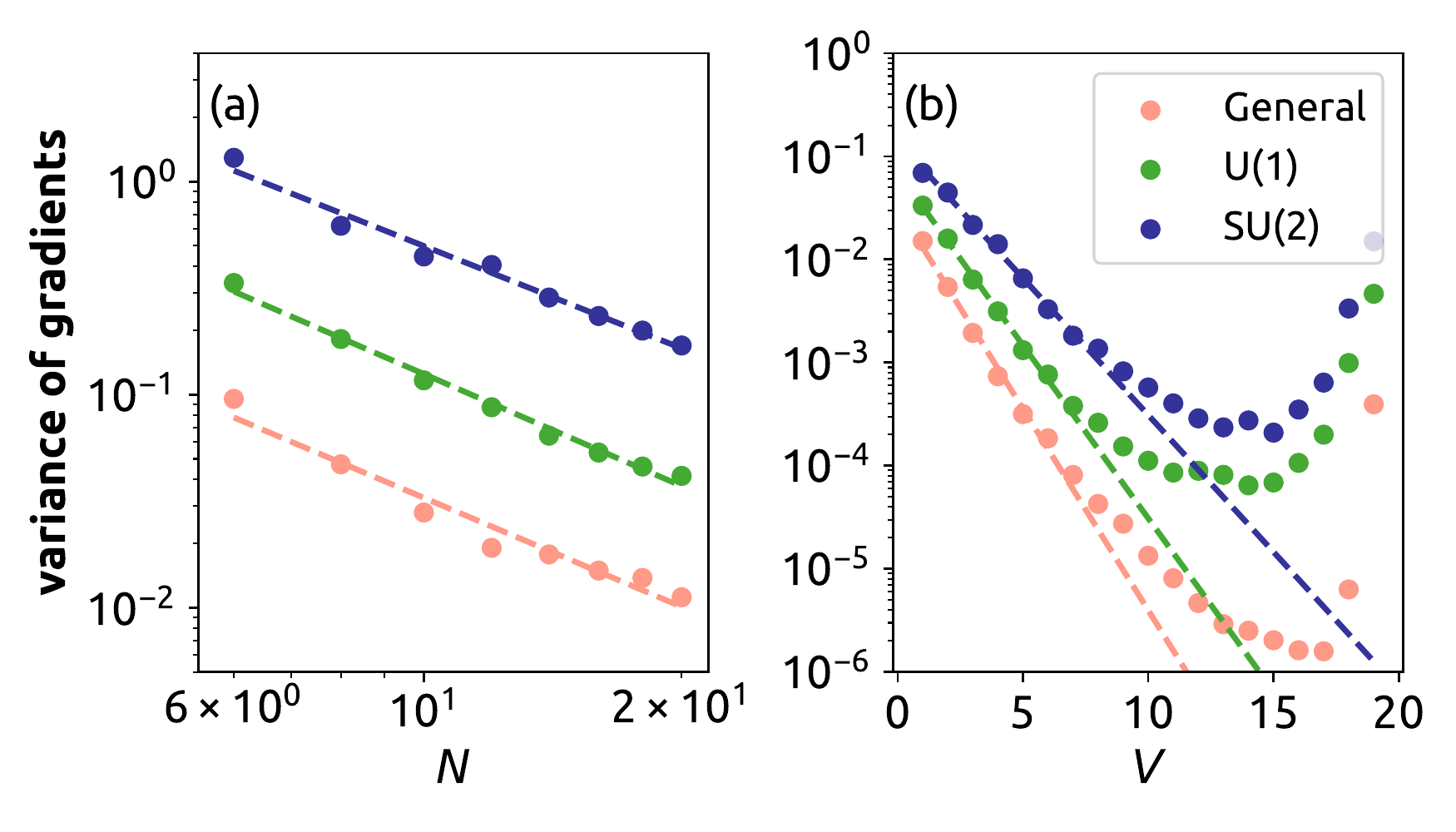}
    \end{center}
    \caption{The variance of the gradient as a function of (a) system size $N$ and (b) virtual qubits number $V$ for random initialized circuit parameters of structures shown in \Fig{fig-circuits}. The Hamiltonian is an open Heisenberg chain. The dashed lines are linear fits, whereas in (b) we only use $V\leq 6$ for fitting.}
    \label{fig-gradvanish}
\end{figure}

First, we consider an unfrustrated Heisenberg model on an open chain of length $N$. Here $N$ should be regarded as the effective circuit width since the output bit strings lie in the Hilbert space of size $2^N$. \Fig{fig-gradvanish}(a) shows the variance of the gradient for circuit blocks with $\UU(1)$ and $\SU(2)$ symmetries with $V=4$.
Interestingly, the variance of the gradient shows a power-law decay in contrast to the exponential decay found in a circuit with generic structure~\cite{McClean2018}. Therefore, it appears that the Q-MPS structure alleviates the gradient vanishing problem at least for the problem under consideration. We attribute this to the fact that the low entropy variational ansatz captures the right inductive bias for the ground state of the target problem.

Next, for an $N=20$ Heisenberg chain, we examine the scaling of the gradient variance with the number of virtual qubits $V$. Again, we see that using symmetry greatly enhance the gradient in \Fig{fig-gradvanish}(b).
We observe an exponential decrease of the gradient in the regime $V\ll N$.
While the gradient increases with $V$ for $V\lesssim N$. Their values are still much smaller than the values in the small $V$ limit, which shows that the Q-MPS are easier to train compared to an unstructured quantum circuit of a generic structure.

Exponential decreasing gradient with respect to $V$ also warns us that there is no free lunch. Although a quantum circuit is able to represent an MPS with exponentially large bond dimension, the number of parameters in the ansatz should also scale exponentially in order to compensate the vanishing gradient. What's worse, the sampling error can smear out too small gradient signals. To really make the ansatz scalable for a highly entangled two dimensional system, we need a parameter efficient design.

\subsection{PEPS inspired ansatz}\label{sec-qpeps}
PEPS is much more parameter efficient for representing the ground states of a two-dimensional quantum lattice Hamiltonians. It is able to represent many area law entangled high dimensional states with a polynomial number of parameters with respect to the system size. However, there is no polynomial time algorithm to contract a PEPS exactly for energy expectation values. Moreover, there is no efficient scheme to prepare a Q-PEPS on a quantum circuit. This task is probably impossible due to the computational complexity argument~\cite{Haferkamp2018}, otherwise, a quantum computer would be able to solve the $\#P$-hard problems~\cite{Biamonte2015,Johnson2013,Biamonte2017}. 
Nevertheless, it is possible to design a variational ansatz that shares the appealing properties of PEPS, such as the 2D area law entanglement entropy.

\begin{figure*}
    \begin{center}
        \includegraphics[width=0.9\textwidth,trim={2cm 3cm 1cm 2cm},clip]{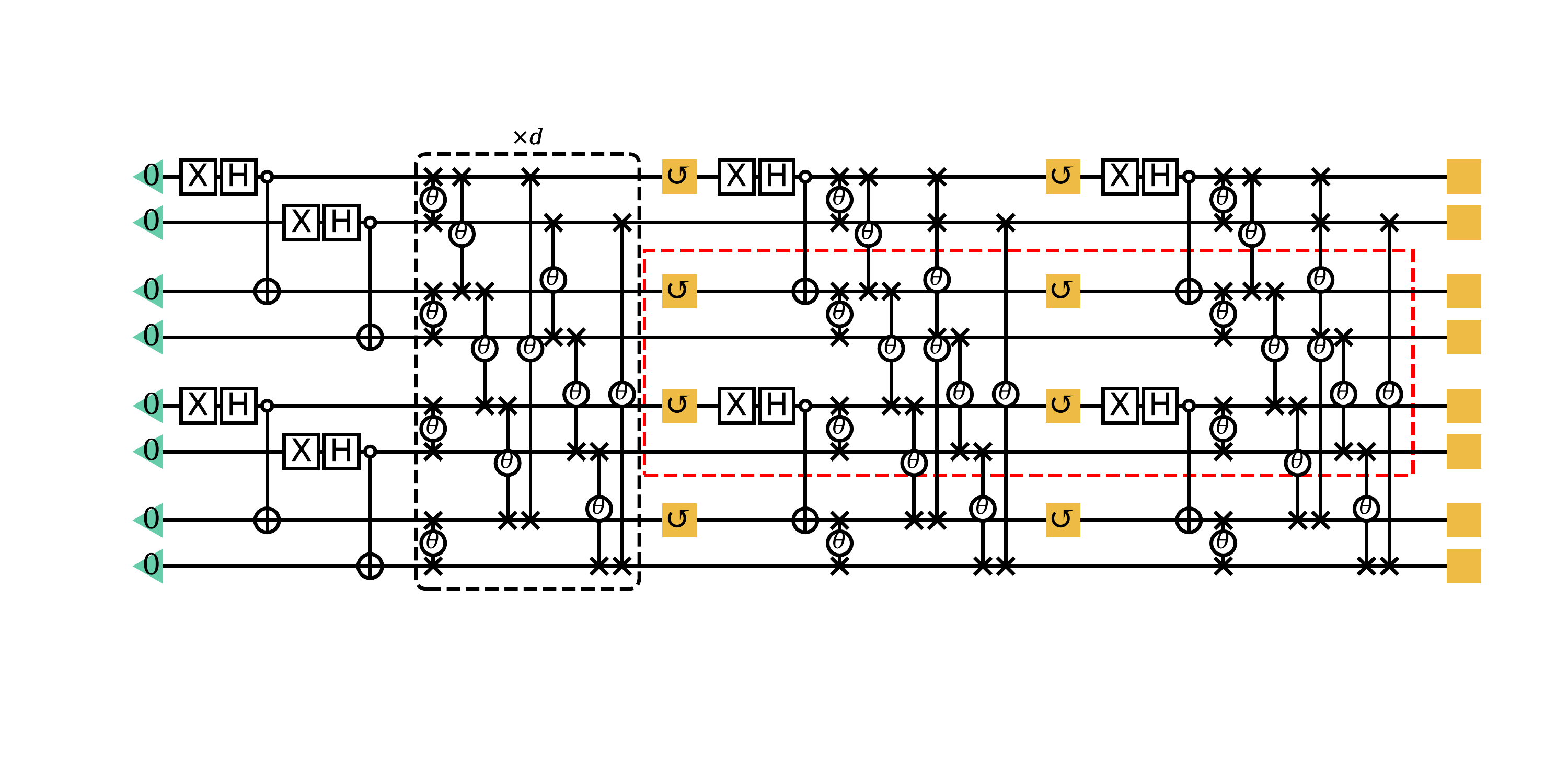}
    \end{center}
    \caption{A projected pair of states inspired variational quantum circuit ansatz with a $4\times4$ square lattice layout.
    Yellow squares with "$\circlearrowleft$" sign are measure (on basis $\alpha=X, Y, Z$) and reset operations. The other yellow squares are measurements without reset.}
    \label{fig-qpeps-circuit}
\end{figure*}

The SU(2) symmetric Q-PEPS ansatz for a $4\times 4$ square lattice is shown in \Fig{fig-qpeps-circuit}.
The entanglement in this ansatz satisfies two-dimensional area law, which can be seen from the red dashed box enclosing a $2\times 2$ region, where the number of bonds crossing the boundary is proportional to the circumference of the box.
In the entangle layer enclosed in the black dashed box, we first entangle each physical qubit with its own virtual qubit(s), then neighboring (periodic boundary condition) physical qubits, and finally neighboring virtual qubits. This completes a single entangle layer. We repeat this layer for $d$ times to increase the number of trainable parameters.
Unlike the case in Q-MPS, where we do not distinguish between virtual qubits, here we assign a constant number of virtual qubits for each physical qubit so that $V/R$ stay as a constant as the system size grows. These physical qubits can be measured in parallel instead of one by one.

\begin{figure}
    \begin{center}
        \includegraphics[width=\columnwidth,trim={0.3cm 0.35cm 0cm 0cm},clip]{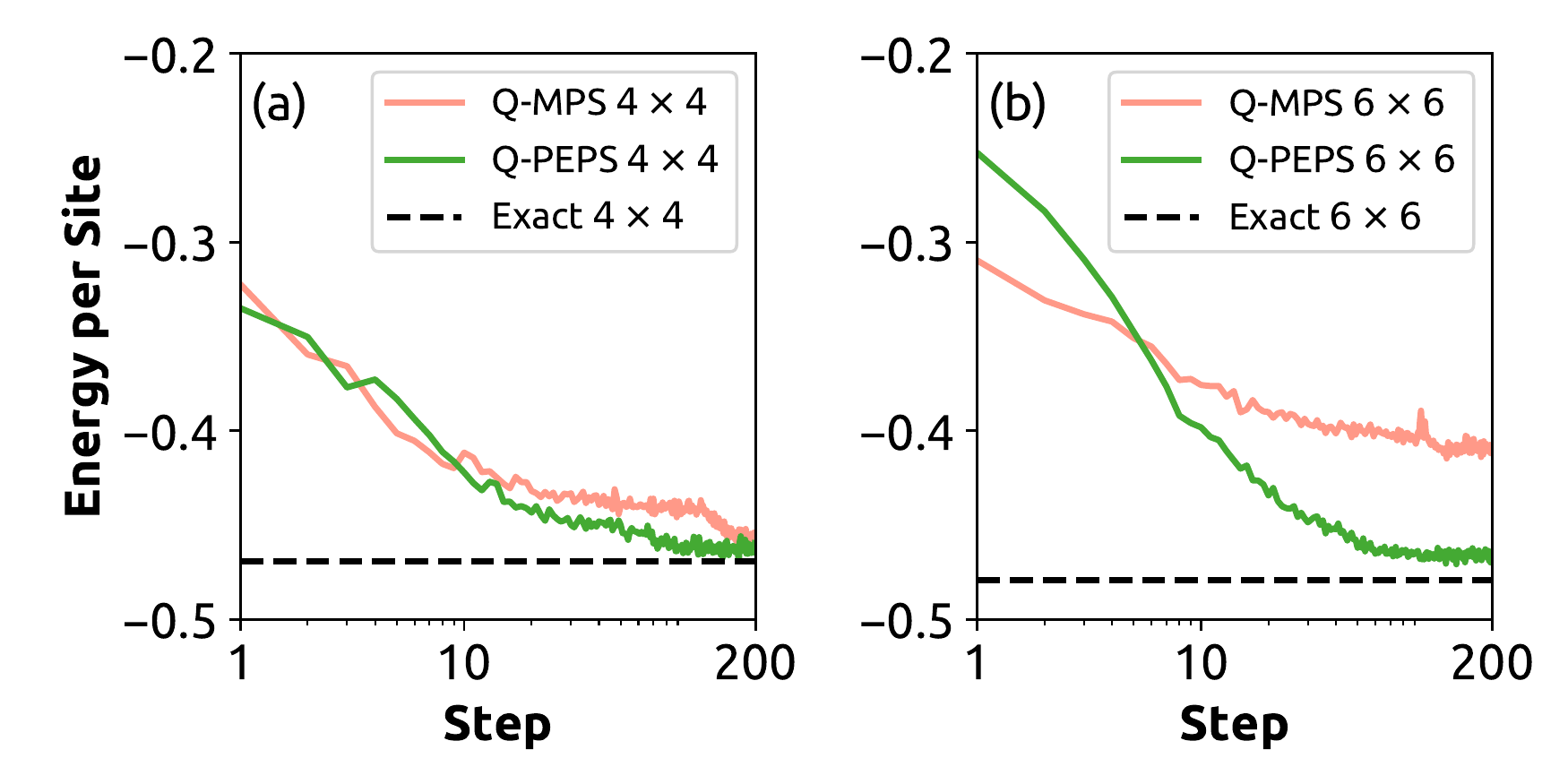}
    \end{center}
    \caption{Energy as a function of training steps, where training parameters are listed in \Tbl{tbl-qpeps}}
    \label{fig-qpeps}
\end{figure}
The simulation results reported in \Fig{fig-qpeps} demonstrates the scalability of this structure on a 2D lattice. In the simulation, we fix the ratio $V/R$ to $1$ and depth $d$ to $5$. So the number of parameters contained in a $4\times 4$ Q-PEPS is $180$, which is the same as that in a depth $d=3$ Q-MPS. The comparative study of two models in \Fig{fig-qpeps} (a) shows similar performance (or slightly better). However, when we scale up the lattice size to $6\times 6$, the ground state energy obtained by Q-PEPS is much lower, indicating better scalability in solving two-dimensional lattice Hamiltonians. One should notice that by increasing the lattice size from $4\times 4$ to a $6 \times 6$, the size of full Hilbert space is increased by a factor of $2^{20}$.

The batch size in the above simulation is $B= 1024$, as the it grows, the  standard deviation of measured gradients $\sigma_s$ will decrease as $\sim 1/\sqrt{B}$. We can also define the variance of gradients as
\begin{equation}
    \sigma_g^2  = \var{\frac{\partial \langle H\rangle}{\partial \theta_i}}{\theta_i},
\end{equation}
where circuit parameters $\thetav_i$ is randomly sampled. If this values is too small, then the gradient vanishes. In a momentum based classical optimizer, the learning rate can be atomatically adjusted to utilize small gradients. However, what matters here is not the absolute value of gradients, but the ratio between the standard deviation of gradients $r_{\rm gs} = \sigma_g/\sigma_s$. When $r_{\rm gs} \ll 1$, the gradient information will be smeared out and the training will probably fail, then we have to either increase the number of shots $B$ or change the ansatz.
Scalable training relies on how does $r_{\rm gs}$ vary as the system size grows.
If we fix the batch size to $4096$, at a system size $4\times 4$, $r_{gs} = 0.142/0.021$, while at a system size $6\times 6$, it is $0.088/0.032$, i.e. the gradient is decreasing and sampling error is increasing. Due to the limit of classical computational power, it requires a quantum device to verify the scalability and fully exploits its power. 
\begin{table}[]\centering
\begin{minipage}{\columnwidth}
\ra{1.3}
    \scalebox{1.1}{
        \begin{tabularx}{0.9\textwidth}{X X X X X}\toprule
            \textbf{Lattice} & $4\times 4$ & $4\times 4$ & $6\times6$ & $6\times6$ \\
            \textbf{Model} & Q-MPS & Q-PEPS& Q-MPS & Q-PEPS\\
            \hline
            \textbf{Gates} & 180 & 180 & 420 & 450\\
            \textbf{Depth $d$} & 3 & 5 & 2 & 5\\
            \textbf{$R+V$} & 6 & 8 & 8 & 12\\
            \textbf{$E/N$} & -0.4596 & -0.4681 & -0.4147 & -0.4707\\
            \textbf{$E_G/N$} & -0.4691 & -0.4691 & -0.4791 & -0.4791\\
            \bottomrule
        \end{tabularx}
    }
    \caption{Parameters used in the numerical experiment in \Fig{fig-qpeps}. The number of shots used in sampling is $1024$, the optimizer is Adam with learning rate $0.1$. $E$ and $E_G$ are the lowest ground state energy obtained in training and exact ground state energy respectively.}\label{tbl-qpeps}
\end{minipage}
\end{table}

\section{Complexity analysis by Gate Counting}\label{sec-gatecount}

\algrenewcommand{\alglinenumber}[1]{\color{red!60}\footnotesize#1:}
\newcommand{\algrule}[1][.2pt]{\par\vskip.5\baselineskip\hrule height #1\par\vskip.5\baselineskip}
\begin{table}
\begin{minipage}{\columnwidth}
\algrule[1pt]
\begin{algorithmic}[1]
    \State $\quad 500$ \Comment{training iterations}
    \State $\times ~M$ \Comment{for each variational parameter}
    \State $\times ~4096$ \Comment{batch size for sampling}
    \State $\times ~2$ \Comment{gradient evaluation}
    \State $\times ~3$ \Comment{Pauli bases}
    \State $\times ~M$ \Comment{parametrized gates}
    \State $\times ~T_\mathrm{gate}$ \Comment{time for each gate}
    \State $\approx 10^7 M^2 T_\mathrm{gate}$.
\end{algorithmic}
\algrule[1pt]
\caption{Gate count for training an Q-MPS with $M$ parameters.}\label{tbl-gatecount}
\end{minipage}
\end{table}

The wall clock time of both classical simulation and actual experiment can be estimated by counting the gate operations, as summarized in \Tbl{tbl-gatecount}. In the classical simulation, the operation time $T_\mathrm{gate}$ scales exponentially with the number of virtual qubits $V$. We employ the differentiable quantum simulator \texttt{Yao.jl}~\cite{Yao} and its GPU backend \texttt{CuYao.jl}~\cite{CuYao} in order to boost the classical simulation efficiency. With native CUDA programming support~\cite{Bezanson2012, Besard2018}, one can parallelize the circuit simulation both for the Hilbert space (of size $2^{V+1}$) and for the batch dimension (of size 4096 in our case). It takes $11-48$ hours (varies for various blocks shown in Fig.~\ref{fig-circuits})  for a typical parameter reported in this letter, $V=4, d=5$ and $N=16$, on a single Nvidia Titan V GPU card.

\begin{table}[]\centering
\begin{minipage}{\columnwidth}
\ra{1.3}
    \scalebox{1.1}{
        \begin{tabularx}{0.9\textwidth}{X X X X X X X X}\toprule
        \textbf{Depth} & 1 & 2 & 3 & 4 & 5 & Exact\\
        \hline
        \textbf{Gates} & 60 & 120 & 180 & 240 & 300 & --\\
        \textbf{Energy} & -0.454 & -0.458 & -0.463 & -0.464 & -0.463 & -0.469\\
        \textbf{Fidelity} & 0.917 & 0.923 & 0.968 & 0.971 & 0.968 & 1.0 \\
    \bottomrule
    \end{tabularx}
    }
    \caption{Variational energy expectation and fidelity of frustrated Heisenberg model with $J_2=0.5$ for various block depth $d$ shown in \Fig{fig-circuits}(c). We only count the number of parametrized gates in the table.
    }\label{tbl-depth}
\end{minipage}
\end{table}

The gate time of an actual quantum device shows constant scaling with respect to the qubit number.
For experiment on SQUID qubits, $T_\mathrm{gate}\approx25ns$~\cite{Chow2010} for a $\SWAP^\alpha$ gate. Thus, solving the $4\times 4$ Heisenberg model with circuit depth $d=1$ and the same hyperparameters as our numerical experiments will take approximately $18$ minutes on a single quantum processing unit (QPU)  without considering readout time. On the other hand, the typical gate time for ion traps is $T_\mathrm{gate}\approx 0.5\mu s-250\mu s$~\cite{Linke2017, Schafer2018}, which means longer time is required for solving the same model.

Furthermore, we note that gate operations in the lines $2-5$ of \Tbl{tbl-gatecount} are also trivially parallelizable on QPUs. Therefore, we envision that building a cluster of QPUs~\cite{Wecker2015a, Wecker2015b} may provide further advantage for high-throughput gradient estimation of the VQE calculation.  In this case, one only needs classical communications to collect the gradients measured on all QPUs. Technically, having intermediate scale quantum circuits running in parallel is also easier than building a fully entangled large scale quantum computer. In this way, we expect running the variational algorithm on parallel QPUs will soon win over classical processors with exponential gate time.

A practical issue is that the total running time $M T_\mathrm{gate}$ should be less than the coherence time of the qubits, which limits the block depth of the variational ansatz to be shallow circuits on near-term devices. \Tbl{tbl-depth} shows the variational energy and fidelity for various depths $d$ obtained for the $N=4\times 4$ frustrated Heisenberg model at $J_{2}=0.5$. One sees that it is possible to reach fidelity $0.917$ with only $60$ parametrized gates, which is within the reach of the current day quantum technology. For larger problem size, one will need to increase the circuit depth linearly with $N$. Assuming area law entanglement entropy scaling of the system, one also needs to scale the circuit width $V$ linearly with the boundary size for an accurate variational description of the ground state.

A crucial step for the qubit efficient Q-MPS preparation scheme is the measure and reset operation.
The cost of this step is device dependent. It is straightforward for trapped ions. However, for SQUIDs, a single qubit measurement and reset can take several microseconds, which is even slower than applying a gate. Fabricating low-latency quantum circuits that support fast measure and control is a rewarding direction in light of the proposed qubit efficient VQE scheme.
Alternatively, one can employ the same circuit architecture without reusing the measured qubits. In this case, one still has the benefit of enhanced gradient signal.

\section{Discussions}\label{sec-discussion}
Classical quantum many-body computation approaches provide valuable insights to quantum algorithms. Using the proposed qubit efficient VQE scheme, one can access the ground state properties of quantum systems using fewer qubits than the system size. Moreover, by exploiting the physical symmetries in the quantum circuit architecture design, one can alleviate the gradient vanishing problem and speed up the convergence to the ground state.


The use of conserved quantum number in circuit construction also allows one to have access to excited states in various quantum number sectors. With this regard, it is interesting to consider what are the universal gate sets with respect to various physical symmetry constraints. In addition to internal symmetries,  the spatial translational symmetry may be taken into account via parameter sharing in the circuit blocks. Then, it naturally raises the question of whether one can study infinite large periodic systems with a finite number of qubits.

Another interesting direction is to perform time evolution or measure time-dependent quantities in the qubit efficient scheme. Since one does not have access to the full wave function directly in the qubit efficient scheme, the Trotter decomposition based time evolution~\cite{Trotter1959, Lloyd1996} may not be directly applicable. Variational quantum algorithms for time evolution~\cite{Li2017b,Mitarai2019} appears to be a good candidate for this purpose.

\begin{acknowledgments}
We thank Miles Stoudenmire for sharing the idea of Ref.~\cite{Huggins2019} prior to its publication.
We thank Yun-Fei Pu and Ding-Shun Lv for providing valuable information on experimental feasibility.
We thank Norbert Schuch for comment on the $\UU(1)$ preserving circuit construction and Yan-Xia Liu for helpful discussions on Bethe Ansatz. We thank Pan Zhang for generous allocation of GPU hours and Xiu-Zhe Luo for contribution to \texttt{Yao.jl}~\cite{Yao}.
The authors are supported by the National Natural Science Foundation of China under the Grant No.~11774398, the Strategic Priority Research Program of Chinese Academy of Sciences Grant No.~XDB28000000 and the research funding from Huawei Technologies under the Grant No.~YBN2018095185.
\end{acknowledgments}

\bibliographystyle{apsrev4-1}
\bibliography{qmps.bib}
\newpage
\appendix
\section{Qubit efficient scheme for cluster state}\label{app-cluster}

\begin{figure}
    \begin{center}
        \includegraphics[width=\columnwidth,trim={2cm 1cm 0.4cm 0.2cm},clip]{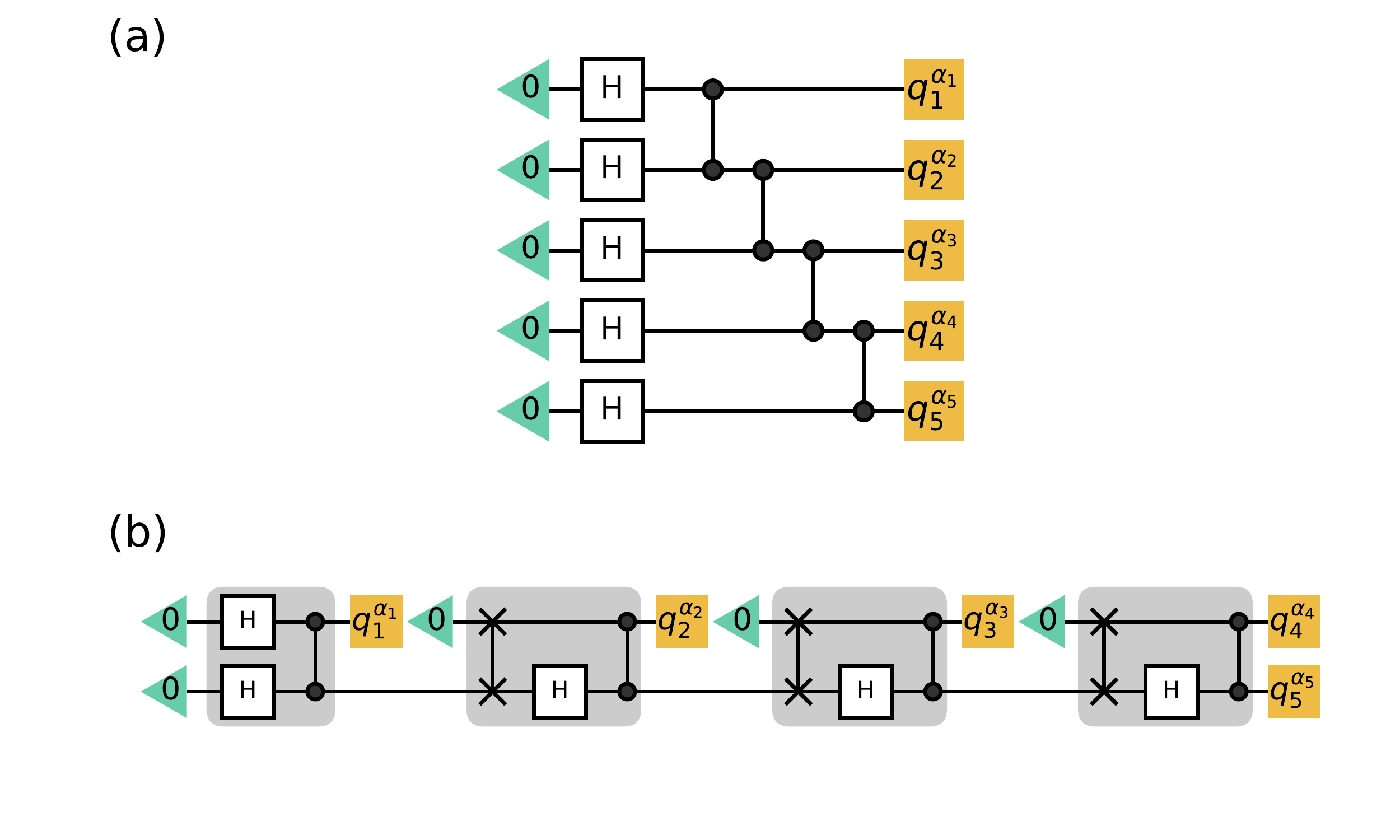}
    \end{center}
    \caption{
    (a) Circuits for preparing and measuring on a five-qubit cluster state. (b) Preparing and measuring the same state in the qubit efficient scheme using only two qubits.
    }
    \label{fig-measure}
\end{figure}

We provide a concrete example of preparing and sampling a cluster state in the qubit efficient scheme.
The one-dimensional (1D) cluster state~\cite{Schon2007} can be prepared by circuit shown in \Fig{fig-measure}.  One first prepares the $N$ qubits to $\ket{+}$ by applying Hadamard gate on $\ket{\mathbf{0}}$ and then sequentially applies CZ gate between the nearest neighbor qubits. Thus, $N$-qubit 1D cluster state can be written as
\begin{equation}\label{eq:cluster_state}
    \ket{\psi_{cl}} = \prod_{i = 1}^{N - 1} CZ_{i, i + 1} \ket{+}^{\otimes N}.
\end{equation}
This 1D cluster state has an equivalent MPS representation with bond dimension two~\cite{d2014quantum}
\begin{equation}\label{eq:cluster_MPS}
\ket{\psi_{cl}} = \sum_{i_1 i_2 \cdots i_N} \tr{A^{(1)}_{i_1} A^{(2)}_{i_2} \cdots A^{(N)}_{i_N}}\ket{i_1 i_2 \cdots i_N},
\end{equation}
where the left and right boundaries are $A^{(1)}_0 = \frac{1}{\sqrt{2}}\bra{+}, A^{(1)}_1 = \frac{1}{\sqrt{2}}\bra{-}$ and $A^{(N)}_0 = \ket{0}, A^{(N)}_1 = \ket{1}$; while for the remaining ones $A^{(i)}_0 = \frac{1}{\sqrt{2}}H, A^{(i)}_1 = \frac{1}{\sqrt{2}}HZ\ (2 \leq i \leq N - 1)$.

One can prepare the same $N$ qubit 1D cluster state using only two qubits. The key is to mediate the entanglement between the physical qubits in different steps using one virtual qubit. As shown in \Fig{fig-measure} (b) we initialize both qubits to $\ket{0}\otimes \ket{0}$, then apply Hadamard gates to create a $\ket{+}\otimes\ket{+}$ state. Applying CZ gates on them we can measure the physical qubit. We then reset the physical qubit to $\ket{0}$ and reuse it. The SWAP gate then exchanges its state with the virtual qubit. We then apply the Hardmard and CZ gates and measure the physical qubits. Repeating these steps to the end, we will obtain the same statistics of the measured bits as in the case of $N$ physical qubits~\Fig{fig-measure}(a).

All two-point Pauli correlation functions vanish on the 1D cluster state. The non-zero correlation functions will emerge if one performs the stochastic local operations and classical communication (SLOCC)~\cite{dur2000three}, e.g., measuring the $i$-th qubit in $Z$ basis and only considering the states with measurement outcome 0. In the general case, for an $N$-qubit 1D cluster state with an SLOCC operator $S$ ($S = DHR_z(\gamma), D = \cos\theta\ket{0}\bra{0} + \sin\theta\ket{1}\bra{1}$) acting on $i$-th qubit, second-nearest-neighbour correlation function $\langle \sigma_{i - 1}^z \sigma_{i + 1}^z \rangle = \cos2\theta\sin\gamma$~\cite{d2014quantum}, and other two-point correlations vanish. In order to measure the correlation function, e.g., $\langle \sigma_2^z \sigma_4^z \rangle$, the estimation protocol is: $\circled{1}$ measuring the $2nd$ and $4th$ qubits in $Z$ basis; $\circled{2}$  performing $HR_z(\pi/2)$ operation on the $3rd$ qubit, measuring it in $Z$ basis and only keeping the states with measurement outcome 0; $\circled{3}$ the measuring bases of the remaining qubits are randomly chosen for the final estimation. Repeat the above procedure until a good estimation of $\langle \sigma_2^z \sigma_4^z \rangle \approx \expect{\sigma_2^z \sigma_4^z}{}$ is obtained.

\end{document}